\documentclass[]{emulateapj}
\newcommand\icarus{Icarus}

\newcommand{\Rsun}{\ensuremath{\mathrm{R}_\odot}}

\newcommand*{\dif}{\mathrm{d}}

\newcommand{\kmps}{\ensuremath{\mathrm{km\ s}^{-1}}\ }

\newcommand{\pdegsqrd}{\ensuremath{\mathrm{deg}^{-2}}\ }

\shorttitle{A search for sub-km KBOs}
\shortauthors{Bickerton et al.}

\begin{document}

\title{A Search for sub-km KBOs with the Method of Serendipitous
  Stellar Occultations}

\author{S.J. Bickerton\altaffilmark{1}}
\affil{Department of Physics \& Astronomy, McMaster University,
  Hamilton, ON L8S 4M1}
\email{bick@physics.mcmaster.ca}

\author{JJ Kavelaars}
\affil{Herzberg Institute of Astrophysics, Victoria, BC V9E 2E7}
\email{JJ.Kavelaars@nrc-cnrc.gc.ca}

\and

\author{D.L. Welch}
\affil{Department of Physics \& Astronomy, McMaster University,
  Hamilton, ON L8S 4M1}
\email{welch@physics.mcmaster.ca}

\altaffiltext{1}{Morton Fellow, Herzberg Institute of Astrophysics,
  Victoria, BC V9E 2E7}

\begin{abstract}
  The results of a search for sub-km Kuiper Belt Objects (KBOs) with
  the method of serendipitous stellar occultations are reported.
  Photometric time series were obtained on the 1.8m telescope at the
  Dominion Astrophysical Observatory (DAO) in Victoria, British
  Columbia, and were analyzed for the presence of occultation events.
  Observations were performed at 40 Hz and included a total of 5.0
  star-hours for target stars in the ecliptic open cluster M35
  ($\beta=0.9\arcdeg$), and 2.1 star-hours for control stars in the
  off-ecliptic open cluster M34 ($\beta=25.7\arcdeg$).  To evaluate
  the recovery fraction of the analysis method, and thereby determine
  the limiting detectable size, artificial occultation events were
  added to simulated time series (1/f scintillation-like
  power-spectra), and to the real data.  No viable candidate
  occultation events were detected.  This limits the cumulative
  surface density of KBOs to $3.5\times10^{10}$ deg$^{-2}$ (95\%
  confidence) for KBOs brighter than m$_R$=35.3 (larger than
  $\sim$860m in diameter, assuming a geometric albedo of 0.04 and a
  distance of 40 AU).  An evaluation of TNO occultations reported in
  the literature suggests that they are unlikely to be genuine, and an
  overall 95\%-confidence upper limit on the surface density of
  $2.8\times10^{9}$ deg$^{-2}$ is obtained for KBOs brighter than
  m$_R$=35 (larger than $\sim$1 km in diameter, assuming a geometric
  albedo of 0.04 and a distance of 40 AU) when all existing surveys
  are combined.
\end{abstract}


\section{Introduction}
\label{sec:intro}

The physical and orbital properties of the Kuiper Belt Objects (KBOs)
are believed to provide valuable information about the formation of
the solar system.  To date, directly observed KBOs include objects
with diameters ranging from $\sim$25km (eg. 1999 DA$_8$,
\citealt{gladman01}; and 2003 BH$_{91}$, \citealt{bernstein04}) to
2400km \citep[eg. Eris,][]{brown06}.  KBOs with diameters below a few
tens of kilometers are beyond the detection of current ground and
space-based telescopes.

The KBO size distribution is the outcome of accretion and collisional
evolution in the outer solar system
\citep{stern96b,davis97,stern97,kenyon99b,kenyon04,pan05}.  The
cumulative luminosity function (CLF) of the KBOs provides an indirect
means of determining their size distribution, and is observed to have
the form: \mbox{$\sum N(<m_R; m_R<25) = 10^{\alpha(m_R-R_0)}$}.  The
KBO size distribution is then a power-law of the form: \mbox{$\dif N
  \propto r^{-q} \dif r$}, with slope $-q$, where $q$ is related to
$\alpha$ via: $q=5\alpha+1$ \citep[][G01]{gladman01}.
Current observations indicate $\alpha\approx0.68$ \citep{fraser07}, or
equivalently, $q=4.6$.  HST observations by \citet{bernstein04}
suggest a lower CLF slope for $m_R>25$.

Current theoretical models of the accretion and fragmentation
processes in the early solar system suggest that $q$ remains constant
only for larger objects, and that the slope of the size distribution
changes at the `break radius', $r_b<$100 km \citep{kenyon99a,
  kenyon99b, kenyon04, pan05}.  Objects smaller than the break radius
would be in collisional equilibrium and their size distribution would
have the so-called Dohnanyi equilibrium slope of $q\approx3.5$
\citep{dohnanyi69}.  Object densities and bulk-moduli can influence
the slope for the small objects, and the position of the break-radius
\citep{kenyon04}.

The deepest direct observations to date \citep{bernstein04} were
performed with HST and reached a limiting magnitude of R=28.5
(corresponding to an object diameter of $\sim$20km for an albedo of
0.04 at 40AU).  The number of observed objects was $\sim$25$\times$
lower than the extrapolation of the current CLF slope.  This absence
of faint objects indicates a decrease in the density of the Kuiper
Belt beyond 50 AU; and a dearth of sub-100km KBOs, in general.  Direct
observation is not a feasible option for pursuing the much fainter
km-sized KBOs.

Some indirect methods of inferring the slope in the range of the
km-sized objects have been explored.  \citet{stern00} used Voyager 2
images of Neptune's moon Triton to analyze cratering and infer a slope
of $q\approx3$ for the size distribution of the km-sized impactors.
\citet{kenyon01} have placed upper limits on small object densities by
examining the optical and FIR sky surface brightnesses of the ecliptic.
They found that a straight extrapolation of the CLF to R$\approx$40-50
mag would yield an ecliptic surface brightness higher than that
observed.

One promising method of indirect detection for the smaller KBOs is
that of serendipitous stellar occultations
\citep{bailey76,dyson92,brown97,roques00,cooray03a,cooray03b,gaudi04,nihei07}.
A km-sized KBO passing through the line of sight to a star will
partially obscure the light from that star to reveal the presence of
the otherwise invisible object.  Such an occultation cannot be
predicted and the practice relies upon serendipity.

A 1 km diameter KBO at 40AU subtends and angle of only 34 $\mu$as, and
the lightcurve resulting from an occultation by such an object is
expected to exhibit a series of increases and decreases due to
diffraction \citep{roques87} (henceforth R87).  At opposition, the
orbital velocities of the Earth (30 \kmps) and a typical KBO at 40 AU
($\sim$4 \kmps) produce a retrograde relative velocity of $v\approx$26
\kmps for the occulter, and the characteristic diffraction shadow cast
by a 1 km KBO will pass over a terrestrial observer in a fraction of a
second.  Observations must be performed at $\geq$40 Hz to sample the
flux oscillation in an observed lightcurve.

We have performed a KBO occultation search with the 1.8m Plaskett
Telescope at the Dominion Astrophysical Observatory (DAO) using a 40Hz
camera.  Models of occultation lightcurves are presented in
Section~\ref{sec:model}.  Observation details are given in
Section~\ref{sec:observations}, with data reduction and analysis
methods described in Sections~\ref{sec:reduction},
and~\ref{sec:detection}, respectively.  Our results are presented in
Section~\ref{sec:results}, and reported KBO occultation detections are
evaluated in Section~\ref{sec:critique}.  A surface density upper
limit is presented in Section~\ref{sec:upperlimit}, with a discussion
and summary provided in Sections~\ref{sec:discussion}
and~\ref{sec:summary}, respectively.

\section{Modeling Occultations}
\label{sec:model}

The object diameter at which diffraction effects begin to dominate a
KBO occultation lightcurve is known as the Fresnel scale, $\rho =
\sqrt{\lambda z/2}$, where $\lambda$ is the wavelength and $z$ is the
distance between the observer and the occulting body.  For a KBO
occultation ($z$=40AU) observed in visible light (550nm), the Fresnel
scale is 1285m.  Fresnel-Kirchhoff diffraction theory should be used
to model occultation lightcurves for objects with sizes of order 1
Fresnel scale unit (Fsu).  Following R87, we have modeled occulters as
circular disks, with the masking disk assembled by placing
progressively smaller rectangular masks at the periphery of the
structure.  Five `orders' of such rectangles were used for our models.
An analytic circularly-symmetric solution to the Fresnel-Kirchhoff
equation exists (R87), but we chose to develop our KBO disks from
assemblies of rectangles to build in flexibility of object shape for
future work.

\subsection{Passbands and Source Dimensions}
\label{sec:nLamb}

\begin{figure}[htbp]
  \plotone{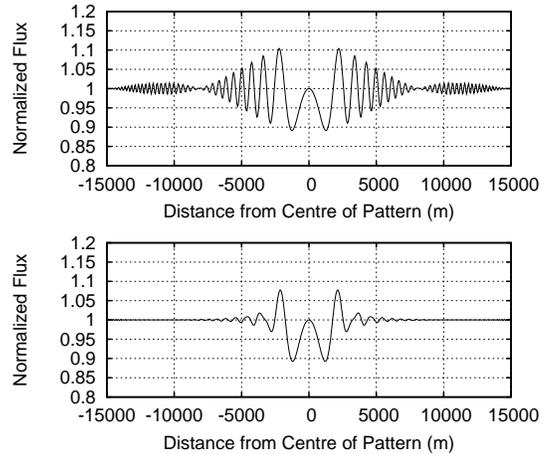}
  \caption[Diffraction profiles in monochromatic and visible light]
  {Diffraction profiles for monochromatic light (550nm) and broadband
    visible light (400-700nm).  The occulting object was 250m in
    radius at a distance of 40AU.}
  \label{fig:monoVsVisible}
\end{figure}

Intensities were numerically integrated over a range of wavelengths to
model a broadband observation of an occultation.  This was
accomplished by averaging evenly-weighted diffraction patterns at
wavelengths distributed uniformly across the passband.  Lightcurves
for 550nm monochromatic light and 400-700nm broadband light are
compared in Figure~\ref{fig:monoVsVisible}.  The extended `ringing'
present in the monochromatic pattern is attenuated by interference in
the broadband pattern; only the first peak away from the pattern
center is well preserved.

It was unnecessary to weight the intensities to match the spectral
energy distribution (SED) of the target star, the spectral throughput
of the telescope, and the spectral sensitivity of the CCD.
Even-weighting offered greater precision than needed for work with
photometric time series having $\gtrsim 1$\% RMS variability.

The target stars were modeled by averaging evenly-spaced point sources
over the projected stellar disks.  A point source density of 100
Fsu$^{-2}$ was used ($\sim$100 points for the stellar disks of the M35
targets projected at 40 AU).  This density was more than sufficient to
provide an accurate model of the shadow, consistent with models found in
\citet{roques00}.

\subsection{Conversion of the Shadow Profile to a Time Series Profile}
\label{sec:shadow2time}

Intensities for the diffraction-dominated occultation shadows were
computed as a function of position with respect to the center of the
projected shadow.  The relative velocity was used to convert
positional coordinates to time coordinates to determine the
corresponding lightcurve that would be recorded by an observer
passing through such a shadow.

\begin{figure}[htbp]
  \centering 
  \includegraphics[viewport= 120 150 410 444,clip=true,width=8cm]{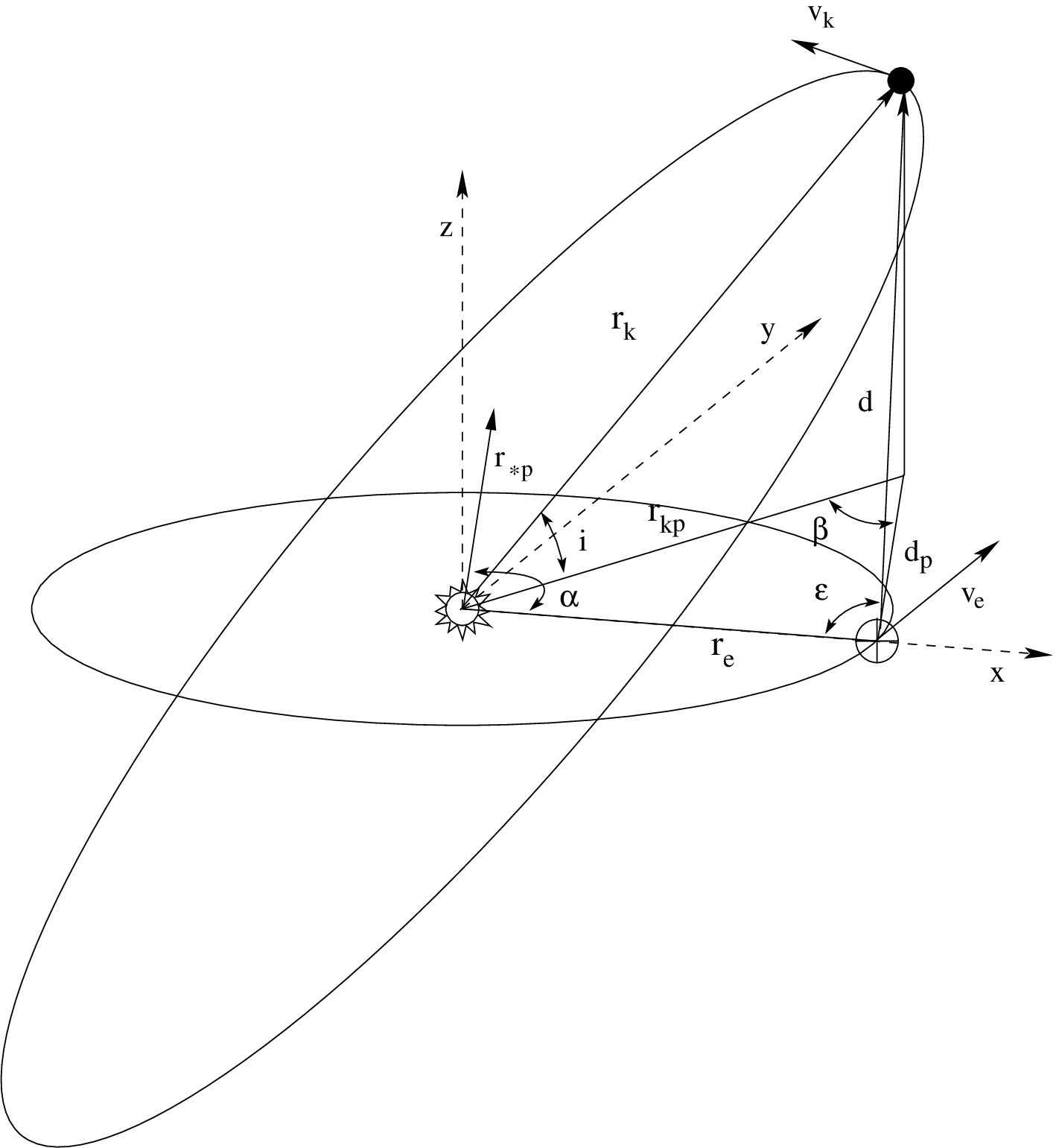}
  \caption[The orbital geometry used to derive the relative velocity
  of a body orbiting on an inclined orbit.]
  {The orbital geometry used to derive the relative velocity of a body
    orbiting on an inclined orbit.  Vectors labeled '$\mathbf{r}$' and
    '$\mathbf{v}$' denote heliocentric distance and velocity,
    respectively.  The vector '$\mathbf{d}$' is the distance from the
    Earth to the occulting KBO.  The subscript 'k' denotes parameters
    referring to a KBO, and 'e' denotes those referring to the Earth.
    The subscript 'p' is used to indicate quantities projected onto
    the ecliptic.  The Earth and KBO are indicated with $\oplus$, and
    $\bullet$ symbols, respectively.  The angle $\varepsilon$ is the
    solar elongation projected onto the ecliptic plane.  The angles
    $\alpha$ and $\beta$ are measured in the ecliptic plane; $\alpha$
    measures the angle between the position vector of the Earth,
    $\mathbf{r}_e$, and the line-of-sight to the target star projected
    onto the ecliptic plane, $\mathbf{r}_{*p}$ ($\mathbf{r}_{*p}$ is
    parallel to $\mathbf{d}_p$). }
  \label{fig:velocityGeo3D}
\end{figure}

An object observed at a given ecliptic latitude must have an orbital
inclination equal to, or greater than, that latitude. As the fraction
of TNOs is observed to decrease with increasing orbital inclination
\citep{jewitt96,brown01}; an observed object is most probably very
near its highest latitude, where its velocity vector lies nearly
parallel to the ecliptic plane.  As only the ecliptic latitude of
observation is known, the orbital inclination was taken to be
{\itshape equal} to the ecliptic latitude.  This is the most probable
configuration, and it is illustrated in
Figure~\ref{fig:velocityGeo3D}.

The relevant velocity is the component of the relative velocity
projected onto the plane perpendicular to the line-of-sight.  Letting
$\theta$ represent the angle between the line-of-sight, $\vec{d}$, and
the relative velocity vector, $\vec{v_{rel}}$, that component is
\mbox{$v_{rel-perp} = |\vec{v_{rel}}| \sin(\theta)$}.  Calculation of
$\theta$ can be performed with the dot product $\vec{d} \cdot
\vec{v_{rel}}$:

\begin{equation}
  \label{eq:dotprod}
  \theta = \arccos \left(\frac{\mathbf{d}\cdot\mathbf{v}_{rel}}{ |\mathbf{d}| |\mathbf{v}_{rel}| }\right).
\end{equation}

\noindent The vectors $\mathbf{d}$ and $\mathbf{v}_{rel}$ were expressed in
Cartesian coordinates in terms of the quantities displayed in
Figure~\ref{fig:velocityGeo3D}. The quantities: $r_{kp}$ (heliocentric
distance to the KBO projected onto the ecliptic place); angles $\beta$,
and $\alpha$; and $d_p$ (observer-to-KBO distance projected onto the
ecliptic) were determined.

\begin{eqnarray}
  \label{eq:rkp}
|\mathbf{r}_{kp}| =&\ |\mathbf{r}_k| \ \cos(i),\\
\beta  =&\ \sin^{-1} \left( \frac{ \sin(\varepsilon)}{ |\mathbf{r}_{kp}| }
\right),\\
\alpha =&\ \pi - \varepsilon,\\
|\mathbf{d}_p| =&\ \sqrt{|\mathbf{r}_{kp}|^2 + |\mathbf{r}_e|^2 - 2\ |\mathbf{r}_{kp}|\ |\mathbf{r}_e|\ \cos(\alpha - \beta)}.
\end{eqnarray}

\noindent The position coordinates are:

\begin{eqnarray}
  (x_e,\ y_e,\ z_e) =&\ \ (|\mathbf{r}_e|,\ 0,\ 0), \quad \mathrm{and}\\
  (x_k,\ y_k,\ z_k) =& 
  \left( \begin{array}{c} |\mathbf{r}_{kp}| \cos(\alpha - \beta)\\
      |\mathbf{r}_{kp}| \sin(\alpha - \beta)\\
      |\mathbf{r}_k| \sin(i) \end{array} \right).
\end{eqnarray}

\noindent The velocity components are:

\begin{eqnarray}
  ( v_{xe},\ v_{ye},\ v_{ze} ) =& (0,\ |\mathbf{v}_e|,\ 0), \quad
  \mathrm{and}\\ 
  ( v_{xk},\ v_{yk},\ v_{zk} ) =& 
  \left(  \begin{array}{c}
      -|\mathbf{v}_k| \sin(\alpha - \beta)\\
      |\mathbf{v}_k| \cos(\alpha - \beta)\\
      0 
      \end{array}\right).
\end{eqnarray}

\noindent The vectors $\mathbf{d}$, and $\mathbf{v}_{rel}$ are then:

\begin{eqnarray}
  \label{eq:vecd}
  \mathbf{d} =& ( x_k-x_e,\ y_k-y_e,\ z_k-z_e ),
  \quad \mathrm{and}\\
  \mathbf{v}_{rel} =& ( v_{xk}-v_{xe},\ v_{yk}-v_{ye},\ v_{zk}-v_{ze} )
\end{eqnarray}

Vectors $\mathbf{d}$ and $\mathbf{v}_{rel}$ were substituted into
Equation~\ref{eq:dotprod} to compute $\theta$; and, in turn, the
perpendicular component of the relative velocity
\mbox{$v_{rel-perp} = |\mathbf{v}_{rel}| \sin(\theta)$}.

If $\mathbf{r}_{kp}$ is less than $\mathbf{r}_e$, the algorithm will fail for certain
elongations as they no longer exist.  For example, an object on an
orbit with $\mathbf{r}_{kp} < \mathbf{r}_e$ cannot be observed at opposition; its
projected position is in the same hemisphere as the sun.


\section{Observations}
\label{sec:observations}

The open clusters: M35 ($\beta=0.9\arcdeg$ N), and M34
($\beta=25.7\arcdeg$ N) were chosen as our primary target field, and
our off-ecliptic control field, respectively.  Two adjacent stars were
monitored in each field.  The properties of the targets are listed in
Table~\ref{tab:stars}.  Values of m$_V$, B-V, and the spectral (MK)
types for the M34 stars were taken from \citet{kharchenko04}. Those
for M35 were taken from \citet{sung99}, but the spectral types were
not available and had to be inferred by comparison of M$_V$ and B-V
values to those for other upper main sequence stars of known spectral
type \citep[reported in][]{kharchenko04} in the cluster. Values for the
stellar radii, R$^*$, were estimated based on their spectral types
\citep{cox00}.  The distance moduli \citep{sarajedini04} were used
to compute the distances with \mbox{$(m - M)_V = 5\ \mathrm{log}(d) -
  5$}.  Values of $R^{*}_{\mathrm{40 AU}}$ denote the stellar radii
projected at 40 AU.  Both clusters are sufficiently distant that the
selected stars have projected diameters less than the Fresnel scale
for visible light at 40AU ($\sim$1300m).

\begin{table}[htbp]
  \centering
  \caption[Properties of stars in the target (M35) and control (M34) fields.]
  {Properties of the stars in the target field (M35), and control field (M34).}
  \begin{tabular}{lll}
    \hline\hline
    & \multicolumn{2}{c}{M34 Targets} \\
    \hline
    Position &              East (left) &   West (right) \\
    ID &                    TYC-2853-22-1 & TYC-2853-679-1 \\
    RA$_{\mathrm{J}2000}$&  02:41:58.31 &   02:41:56.78 \\   
    Dec$_{\mathrm{J}2000}$ &+42:47:29.7 &   +42:47:22.4 \\    
    $\beta$                & 25.7$\arcdeg$ N & 25.7$\arcdeg$ N \\
    m$_V$ &                 8.360    &      8.444 \\           
    B-V &                   0.002 &         -0.005 \\        
    MK&          B9V &           B9Vp Mn\\
    R$^\star$ &             3.0 $\Rsun$ &   3.0 $\Rsun$\\
    R$^\star_{40\mathrm{AU}}$ & 640m &      640m\\
    (m-M)$_V$ & 8.98$\pm$0.06 & 8.98$\pm$0.06 \\
    Dist. & 625$\pm$17 pc & 625$\pm$17 pc \\ 
    \hline
    & \multicolumn{2}{c}{M35 Targets} \\
    \hline
    Position &                  East (left) &   West (right) \\ 
    ID &                        TYC-1877-704-2 &TYC-1877-266-1\\
    RA$_{\mathrm{J}2000}$&      06:09:20.75 &   06:09:18.50 \\  
    Dec$_{\mathrm{J}2000}$ &    +24:22:08:6 &   +24:22:05.4 \\  
    $\beta$                & 0.9$\arcdeg$ N & 0.9$\arcdeg$ N \\
    m$_V$ &                     10.656 &        9.764 \\        
    B-V &                       0.178 &         0.167 \\        
    MK&              B9V &           B9V\\
    R$^\star$ &                 3.0 $\Rsun$ &   3.0 $\Rsun$\\
    R$^\star_{40\mathrm{AU}}$ & 360m &          360m\\    
    (m-M)$_V$ & 10.21$\pm$0.12 & 10.21$\pm$0.12 \\
    Dist. & 1101$\pm$60 pc & 1101$\pm$60 pc\\
    \hline\hline
  \end{tabular}
  \label{tab:stars}
\end{table}

Observations were conducted in 2-3\arcsec seeing conditions with the
1.8m Plaskett Telescope at the Dominion Astrophysical Observatory
(DAO) on the UT nights of 2004 January 2, and 2004 January 3.  These
dates were chosen to place the primary target near an elongation of
$\varepsilon=180\arcdeg$, where occultation probability is highest
\citep{roques00}.  On those dates, the solar elongations of M35 and
M34 were $\varepsilon=169\arcdeg$, and $\varepsilon=128\arcdeg$,
respectively.

A Texas Instruments TC253-SPD CCD was operated with a San Diego State
University 2 (SDSU2) controller, and data were binned 8$\times$8, and
sub-rastered down to a smaller region of the chip to enable the CCD to
be read-out at 40Hz. The binned pixel scale was 2.67\arcsec
~pixel$^{-1}$. Images taken early in the observing run were
30$\times$24 pixels, and images taken later were 25$\times$16
(8$\times$8 binned, {\itshape un}processed sizes).  One hundred images
were taken in repeated sequences, and were stored in 3D-FITS format.
Each was followed by a $\sim$0.5s overhead as the controller initiated
the next 100-frame sequence. In total, 357,000 images were obtained
for M34, and 794,000 were obtained for M35.

\section{Data Reduction}
\label{sec:reduction}

\begin{figure}[htbp]
  \plottwo{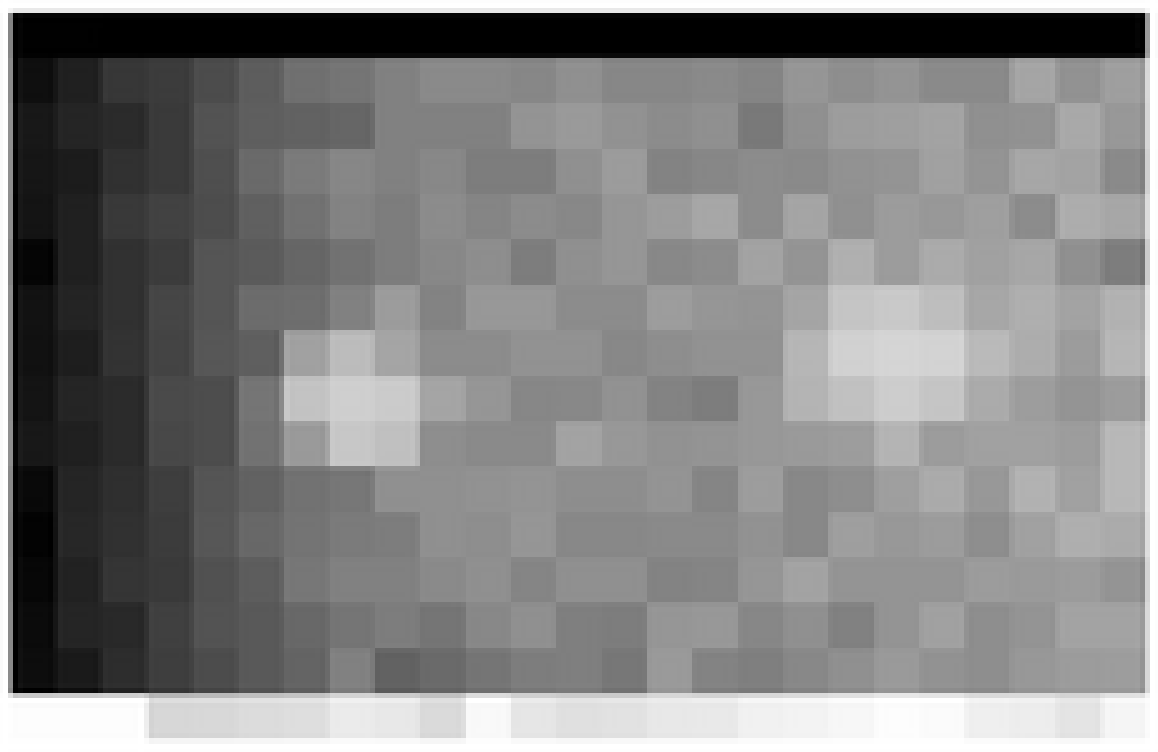}{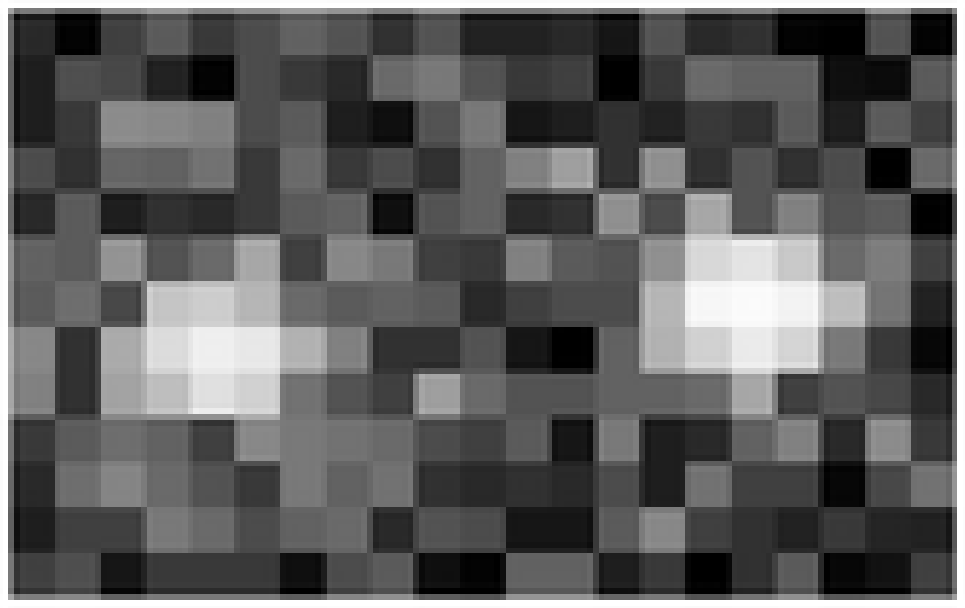}
  \caption[Unprocessed and pre-processed M35 frames.]
  {Unprocessed (left) and a pre-processed (right) M35 frames.
    The frames are 25$\times$16 pixels, and 21$\times$12 pixels,
    respectively. With 8$\times$8 binning, the processed image is
    56$\arcsec \times$32$\arcsec$.}
  \label{fig:m35samples}
\end{figure}

A bias frame was constructed by median-combining 3600 25ms dark
exposures.  The count level was $\sim$3000 ADU with an estimated sky
contribution of $\sim$2 ADU.  One thousand 100ms sky-flat calibration
images were median-combined to produced a flat which showed no
structure and had pixel-to-pixel variability of less than 1\%.  The
true variability across the field was, therefore, assumed to be
negligible, and no flat-field correction was applied.  Examples of raw
and pre-processed images are shown for the M35 field in
Figure~\ref{fig:m35samples}.

As the high-speed camera was installed in place of the guider, guiding
was unavailable during the observations.  Without guiding, the
Plaskett telescope's tracking motor introduced an oscillation in right
ascension with an amplitude of $A\approx8$ arcsec, and a period of
$T\approx$4 minutes.  This corresponds to a maximum drift speed across
the chip of $2\pi A/T\approx$0.2 arcsec s$^{-1}$, or 0.08 pixels
s$^{-1}$.  The target star's motion
between pixels was too small to induce a calibration error on the
sub-second time scale of a KBO occultation.

Photometry was performed with SExtractor v2.3 \citep{bertin96}.  The
stars in the field were not crowded, and aperture photometry was
performed with 5 pixel (13.3 arcsec) apertures (diameter).  Other
SExtractor parameters were set at their default values.  Flux values
for the targets in the 100-frame sequences were concatenated into
longer sequences of $\sim$50,000 -- 100,000 frames.  The 0.5s gaps
occurring between the sequences were masked and ignored.  The gaps are
cosmetically unappealing, but only diminish the usefulness of the data
in proportion to the duty cycle reduction they represent ($\sim17\%$).

Each time series was normalized to have a constant mean flux.  A
sufficiently large number of simultaneously observed targets would
make it possible to normalize by dividing by the summed flux of the
other targets.  With two targets of comparable brightness, this method
would have increased the noise level in a time series by $\sim$40\%
($\sqrt2$).  Each time series was therefore normalized by dividing
by a Gaussian-smoothed version of itself.  To construct smoothed time
series, each point was replaced by the Gaussian-weighted average of
the flux measurements preceding it and following it, such that for
flux $f$, the smoothed flux at the $j^{th}$ time coordinate was:

\begin{equation}
  \label{eq:smooth}
  f_{j} = \frac{\sum_{i} w_{ij} f_i}{\sum_{i} w_{ij}},
\end{equation}

\noindent where $w_{ij}$ is a Gaussian
having a `width' of $2\sigma = 1$s, truncated at
$\pm 3\sigma$.  

\begin{equation}
  \label{eq:kernel}
  w_{ij} = \frac{1}{\sqrt{2\pi}\sigma} e^{\frac{-(x_i-x_j)^2}{2\sigma^2}}.
\end{equation}

Simulated occultation lightcurves with no noise were normalized in the
same way to evaluate the effects of smoothing on the events being
sought.  The amplitudes of these events were found to change by
$<$0.5\%, indicating that the normalization does not significantly
attenuate or suppress an occultation signal.

\begin{figure}[htbp]
  \plotone{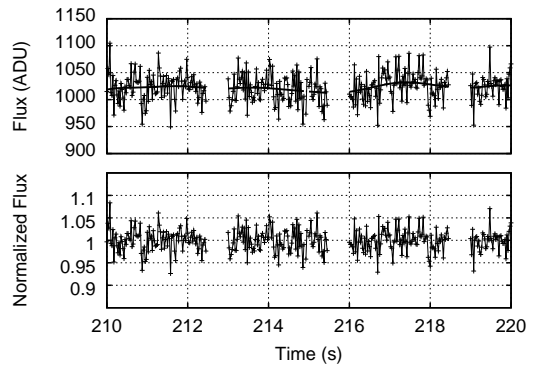}
  \caption[A 10s sample of the time series for eastern star in the
  M35 field.]
  {A 10s sample of the time series for the eastern star in the M35
    field.  Both the raw flux counts (top) and the normalized data
    (bottom) are shown.  The smoothed normalization curve is shown
    superimposed over the raw flux. Gaps mark the dead time of the
    camera.}
  \label{fig:normM35}
\end{figure}

Figure~\ref{fig:normM35} shows a segment of a time series for the
eastern-most star in the M35 field.  The smoothed time series is
superimposed, and the normalized time series is shown in a separate panel.

Imaging continued through periods of reduced transparency,
or partial cloud.  When a star's measured flux dropped by more than
$\sim$30\% in a given time segment, the noise levels in that time
segment increased accordingly by $\sim$15\% ($\sqrt{N}$ photon
statistics), and the segment was masked.

The RMS noise values of the observed flux were computed for each
target.  The variates included all data for a given source; they
represent the overall observation quality.  A 3$\sigma$-clipped
standard deviation was used to avoid the influence of outliers.
The eastern and western stars in the M35
field had noise levels of $\sim$4.3\% and $\sim$3.4\%, respectively.
Both stars in the M34 field had noise levels of $\sim$1.6\%.

\section{The Detection Algorithm}
\label{sec:detection}

An object with a diameter $\gtrsim$1 Fsu (1285m for visible light at
40 AU) will produce a $>$40\% flux decrease during occultation.  No
such events were found and a cross-correlation detection algorithm was
used to search for events caused by sub-Fresnel-scale objects near the
noise limit.

Template diffraction patterns were cross-correlated with each
time series to detect candidate occultation events.  
For a discretely-sampled time series, $t_i$,
and a template `kernel', $k_i$, the cross-correlation is \citep[see
][]{bracewell86}: 

\begin{equation}
  \label{eq:xcorr_discrete}
  (t \star k)_j \equiv \sum_i t_i\ k_{i+j}.
\end{equation}

\noindent A positive peak in $(t \star k)_j$ indicates that $t$ is
similar to $k$ when $k$ is offset to position $j$. The detection
threshold was set to $>+6\sigma_{t\star k}$ (random probability
$\lesssim 10^{-9}$).

Masked data (gaps and high-noise regions) were ignored by the
algorithm, and the total on-source observing time is the number
of non-masked points times the individual exposure time.

To construct detection kernels for cross-correlation, theoretical
occultation lightcurves were integrated and normalized in 0.025s segments to
reproduce the effect of a camera taking exposures of that duration
(our exposure time).  Typically, 8-12 points were sufficient to model
a kernel. 

At 40 Hz sampling, the diffraction features of a KBO occultation
lightcurve are slightly under-sampled.  A small shift in the sample
start time produces subtle changes in the structure of the resulting
kernel, and the cross-correlation could not be performed with kernels
for all possible time-shifts.  To construct {\itshape
  detection}-kernels, the integration start-time was synchronized with
the event such that the kernel produced would be symmetric, and this
single kernel was used as the {\itshape the} detection-kernel.
{\itshape Simulated occultation}-kernels which were planted in a time
series for testing purposes, were integrated with a random start-time.
Simulated occultation events were `planted' in time series to evaluate
the performance and recovery limits of the cross-correlation
algorithm.  For each planted event, a position in the time series was
chosen randomly, and the relevant points in the time series were
multiplied by the corresponding intensities from the kernel.

\subsection{Performance of the Detection Algorithm}
\label{sssec:xcorr}

Artificial time series were used to assess the performance of the
detection algorithm as they were certain to contain no occultations.
Analysis of our time series showed that they were not composed of pure
Poisson noise, but rather had power spectra with slopes of
approximately -1 (due to scintillation).  Random variates were,
therefore, assigned in the frequency domain to produce power spectra
with the appropriate slope, and these were inverse Fourier-transformed
to generate time series \citep{bickerton06}.  This method of generating
artificial time series will be discussed in detail in future
work.

\begin{figure}[htbp]
  \plotone{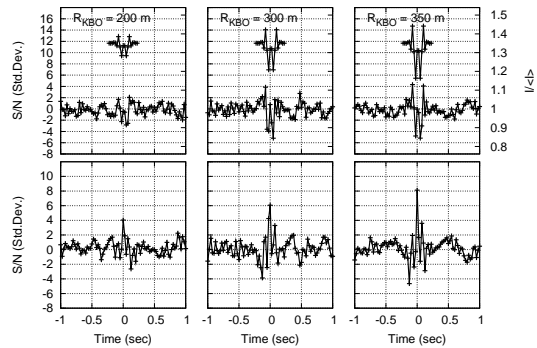}
  \caption[The cross-correlation and low-flux detection methods.] 
  {The cross-correlation and low-flux detection methods.  Planted
    occultation events are shown for 4, 6, and 8$\sigma$
    cross-correlation detections.
    Upper panels show the events in the time series data; the lower
    panels show the cross-correlation values.  The left vertical axes
    are in units of standard deviation from the mean.  The normalized
    flux units are shown on the right vertical axes in the upper
    panels.  Original units for the cross-correlation are not shown.
    SNR values are higher for the cross-correlation variates than in
    the original time series. }
  \label{fig:exampleEvents}
\end{figure}

The signal-to-noise ratio (SNR) of the cross-correlation method was
compared to a `low-flux' method (a search based on identifying
unexpectedly low flux values).  We define `noise' to be the
root-mean-squared (RMS) deviation of the intensities from their mean,
and `signal' to be an event's maximum intensity deviation from the
mean intensity.  Planted occultation events for three different sized
KBOs (at 40 AU) are shown in Figure~\ref{fig:exampleEvents} to
illustrate the advantage of the cross-correlation method.

Any value above the detection threshold was compared to its expected
value before being considered a candidate.  The expected value was
estimated to be the maximum value of the auto-correlation of the
detection kernel ($(k \star k)_j$).  High values in the
cross-correlation were empirically found to be systematically smaller
(0.93$\times$) than the auto-correlation value, and the
auto-correlation value was scaled accordingly.  To cover the range of
possible cross-correlation values, those within $\pm 8\sigma$ (more
than sufficient to include all legitimate occultation candidates) of
the scaled auto-correlation value were accepted as candidates.

The cross-correlation method yielded a higher signal-to-noise than the
low-flux method for all objects tested (r$_{KBO}$=150-500m,
distance=40 AU).  As even a conservative (low)
estimates of the slope of the KBO size distribution is quite steep
\citep[q=3,][]{pan05}, {\itshape the ability to detect objects with
  radii only a few dozen meters smaller could significantly
  increase the total number of objects to which the observations are
  sensitive!}

The time series variates and those of the cross-correlation were
consistent with having been drawn from Gaussian distributions (based
on $\chi^2$ tests).  As we report no positive detections, the issue of
false positive rates is not addressed here.  It will be presented in
future work.

\subsection{Choosing the Detection Kernels}
\label{sssec:kernelset}

The structure of an occultation lightcurve is influenced by the size
of the KBO, the distance to it, and the minimum distance between the
KBO center and that of the stellar disk (called the impact parameter).
Multiple detection kernels were used to provide coverage of the
parameter space.

To select a kernel set, occultation events were planted in artificial
data with a 4\% noise level (the highest mean noise level in our M35
time series).  With the distance and impact parameter held constant,
objects of various sizes were planted and the detection algorithm was
run.  A detection kernel for a single object size was used to
determine the range of object sizes it could successfully detect.
This was repeated with detection kernels for different-sized objects,
and a set with overlapping ranges of sensitivity was selected.  This
process was repeated to determine the ranges of sensitivity in terms
of distance, and impact parameter.

The final kernel set included objects at distances of 10, 20, 40, 80,
and 160 AU, and having sizes of $r_{KBO}$=0.08, 0.16, 0.24, 0.32,
0.40, and 0.48 Fsu (an event generated by an object larger than
$r_{KBO}\gtrsim$0.5 Fsu could be identified by visual examination of
the time series, even with noise levels $>$4\%).  Ten evenly-spaced
impact parameters were included at each size/distance.  Thus, a total
of 300 kernels were cross-correlated with each time series.  The same
kernel set was tested with 2\% noise levels and performed equally
well.

\subsection{The Recovery Limit and Apparent Shadow Size}
\label{ssec:efficiency}

The smallest KBO radius to which the detection method is sensitive,
$r_{min}$, and the maximum impact parameter at which a KBO of this
limiting size could be reliably detected, $b_{max}$, were determined
by a plant and recover process.  Occultation profiles were computed for
the projected stellar radii and photometric passbands of the M34 and
M35 targets, and were planted in artificial time series with noise
levels of 2\% (M34) and 4\% (M35).  The recovery fraction was
determined for a range of KBO radii (20m intervals) and impact parameters
(60m intervals).  

For each radius/impact parameter pair (r,b), 100 events were planted
and the detection algorithm was run.  A given (r,b) pair was
considered detectable if its recovery fraction was $\geq$80\%, and the
detectability limits ($r_{min}$,$b_{max}$) were determined for
occulting KBOs at different distances (see Table~\ref{tab:densitylimit}).

\begin{table}[htbp]
  \centering
  \caption[The limiting TNO surface densities indicated by the null result]
  {The limiting TNO surface densities indicated by the null result.}
  \begin{tabular}{ccc cc}
    \hline\hline

    d & r$_{min}$ &  R & $b_{max}$ & $\sigma_{0}$ (.95)\\

    AU & m & mag & km & \pdegsqrd\\

    \hline

    &&&&\\
    \multicolumn{5}{c}{Off-Ecliptic ($\beta=25.7\arcdeg$) (M34)}\\
    &&&&\\
    10  & 200  & 30.90  & 1.2  &  5.3e+09 \\
    20  & 220  & 33.70  & 1.2  &  1.9e+10 \\
    40  & 320  & 35.90  & 1.5  &  5.9e+10 \\
    80  & 500  & 37.94  & 1.8  &  2.1e+11 \\
    160  & 750  & 40.07  & 1.5  &  1.1e+12 \\
    &&&&\\
    \multicolumn{5}{c}{Ecliptic ($\beta=0.9\arcdeg$) (M35)}\\
    &&&&\\
    10  & 230  & 30.60  & 0.8  &  5.1e+09 \\
    20  & 300  & 33.03  & 1.1  &  1.3e+10 \\
    {\bfseries  40 } &{\bfseries  430 } &{\bfseries  35.26 } &{\bfseries  1.5 } & {\bfseries  3.5e+10 }\\
    80  & 600  & 37.55  & 1.8  &  1.2e+11 \\
    160  & 850  & 39.80  & 2.0  &  4.4e+11 \\
    &&&&\\
    
    \hline\hline
  \end{tabular}
  \label{tab:densitylimit}
\end{table}

\section{Results}
\label{sec:results}

The algorithm flagged 177 events for review: 133 in the M35 data (27
per hour), and 44 in the M34 control data (21 per hour).  To review
each reported detection, the time segment surrounding it ($\pm$1.5s)
was examined and compared with that of the adjacent star, and with the
theoretical lightcurve of the detection-kernel.  Raw, unreduced flux
values of the two time series were also examined to provide an
indication of any artifacts which may have been introduced by the
reduction and normalization processes.

\begin{figure}[htbp]
  \plotone{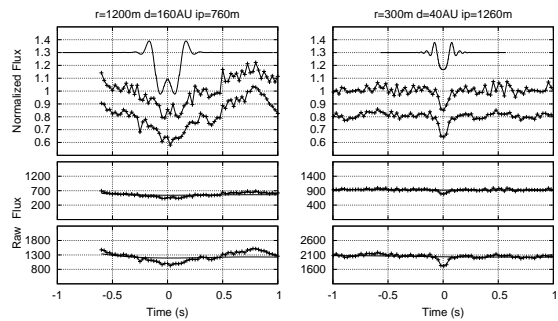}
  \caption[Lightcurves of two false positive events from the M35 data.]
  {Lightcurves of two false positive events from the M35 data.  The
    upper panels show the normalized flux for both stars (flux for the
    neighboring star is offset for clarity). The parameters of the
    kernel which detected the events are printed above each plot
    (r=KBO radius, d=distance, and ip=impact parameter), and the
    corresponding theoretical occultation profiles are shown above the
    stellar lightcurves.  For each event, the raw flux values for the
    two stars are shown in the lower two panels. In each case, the
    coincident flux decrease observed in the neighboring star
    indicates that the event is inconsistent with interpretation as a
    KBO occultation.}
  \label{fig:falsePositives}
\end{figure}

{\itshape\bfseries The cross-correlation detection method and the
  subsequent review process found no viable occultation candidates in
  our time series.}  All flagged photometric fluctuations were clearly
present in both stars, and the events were easily identified as being
spurious.  Variations occurring on time-scales comparable to the
duration of the smoothing kernel (1.0s) were only common in the more
unstable M35 time series, and did not resemble occultations.  Those
occurring on shorter time-scales were common in all time series, and
did in many cases, resemble occultations.  Examples are shown in
Figure~\ref{fig:falsePositives}.

\begin{figure}[htbp]
  \plotone{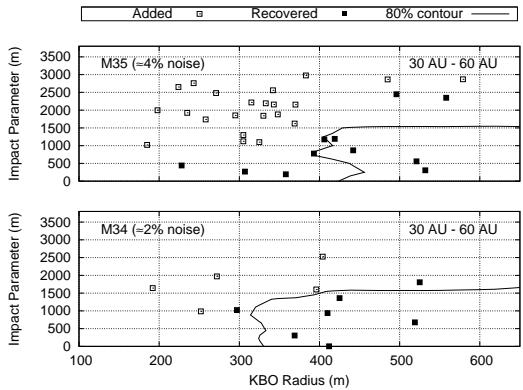}
  \caption[The recovered sizes and impact parameters of planted
  occultation events] 
  {The recovered sizes and impact parameters of planted occultation
    events for 30-60AU objects.  Squares represent planted occultation
    events; filled if recovered, and open if not recovered.  The 80\%
    contour from the theoretical recovery tests for 40AU objects
    (Section~\ref{ssec:efficiency}) is shown with a solid line.}
  \label{fig:recovered_kbos}
\end{figure}

To evaluate the validity of the recovery tests, objects were added to
the real, normalized, time series prior to running the detection
algorithm.  This was done to ensure that the manual review process did
not alter the recovery rates significantly.  The sizes and impact
parameters of the events added and recovered are shown in
Figure~\ref{fig:recovered_kbos}, with the 80\% contour from the 40AU
theoretical recovery tests (Section~\ref{ssec:efficiency}) overlaid
for reference. The theoretical limit shown is based on 40AU tests, and
the planted events shown are those with distances of 30AU $<$ d $<$
60AU.  Some of the planted events are outside the 80\% contour as a
result of the broader distance range.

With this null result, we can place an upper limit on the surface
density of TNOs.  Taking a line of length 1 arcsec perpendicular to
the direction of motion, TNOs of density $\sigma$ arcsec$^{-2}$,
moving at $v$ arcsec s$^{-1}$, will cross the line at a rate $\sigma
v$ s$^{-1}$.  After $t$ seconds, $\sigma v t$ objects will have
crossed the 1 arcsec line.  For a star positioned somewhere on the
line, the probability that $x$ TNO line-crossing events will lie
within 1 impact parameter, $b$, of the star (along the line) is
described by Poisson statistics with the rate $2bvt\sigma$.

\begin{equation}
  \label{eq:linecross}
  P(x; b, v, t, \sigma) = \frac{(2bvt\sigma)^{x}}{x!} e^{-2bvt\sigma}.
\end{equation}

As $\sigma_0$ is assumed constant, only observations expected to have
the same value for $\sigma_0$ (same ecliptic latitude) should be included.

The product $bvt$ represents an area swept through by the target star,
and could be integrated over $b(t)v(t)\dif t$ to account for
variability in $b$ and $v$, or to incorporate results from multiple
stars.  Our values of $b$ and $v$ were similar for both stars, and did
not vary in time.  The time $t$ represented the sum of observing time
on both stars.

The 95\% confidence limits for $\sigma$ occur when the probability of
detection is 0.05 (ie. the complement of equation~\ref{eq:linecross}).
With $x=0$ events, equation~\ref{eq:linecross} can be reduced, and
solved in terms of $\sigma_0$:

\begin{equation}
  \label{eq:densitylimit}
  \sigma_0 = \frac{\ln | 1-P |}{-2 b v t}.
\end{equation}

The limiting densities indicated by our null result are presented in
Table~\ref{tab:densitylimit}.  Values of $r_{min}$ (the size for which
the result is relevant) and $b_{max}$ were determined in the recovery
tests described in Section~\ref{ssec:efficiency}.  The relative
velocities were computed as described in
Section~\ref{sec:shadow2time}, and the time values are the total
observed times (masked data not included).

The limit for typical KBOs (40 AU, $\beta\approx 0$) was placed on the
cumulative luminosity function (CLF) by converting the object size
(r$_{min}$) to an R magnitude (G01) (assumes geometric albedo of 0.04,
subscripts denote units).

\begin{equation}
  \label{eq:size2mag}
  m_R \simeq 25.9 + 2.5 \log \left[ \left(\frac{d_{\mbox{AU}}}{50}\right)^4
      \left(\frac{100}{2r_{\mbox{km}}}\right)^2 \right].
\end{equation}

A radius of 430m corresponds an R magnitude of $\sim$35.3.  Although
the limit is placed on the {\itshape cumulative} luminosity function,
equation~\ref{eq:densitylimit} assumes a uniform object size.  This
assumption is reasonable for even the most conservative (shallow)
estimates of the slope of the size distribution, as they place the
majority of detectable KBOs at the smallest size limit.

\section{Evaluation of Reported TNO Occultations}
\label{sec:critique}

TNO occultations have been reported by two separate groups:
\citet{chang06}, and \citet{roques06}.  The TNO densities implied by
these positive results are surprisingly high; here we evaluate
their methods and results.

\subsection{TNOs detected by \citet{chang06}}

\citet{chang06} (henceforth C06) searched $\sim$89 hours of
time-tagged x-ray photon data from the Rossi X-ray Timing Explorer
(RXTE) archive, and reported 58 TNO occultation events.  Many of the
events were coincident with high-energy particle arrivals recorded by
other instrumentation on-board RXTE \citep{jones06}, and 12
occultation candidates remained after re-analysis \citet{chang07}.
\citet{jones07} report that, at most, 10\% of the observed lightcurve
dips could be due to TNO occultations; and they suggest that it would
be a mistake to conclude that {\itshape any} TNO occultations have
been detected in the Sco X-1 X-ray data.

Our observations were not sensitive to objects in the size range
observed by C06, and there is no disagreement with our null
result. But the C06 result indicates a surface density $\sim 10^6$
times greater than that estimated by extrapolation of the CLF reported
by \citet{bernstein04}.

\begin{figure}[htbp]
  \plotone{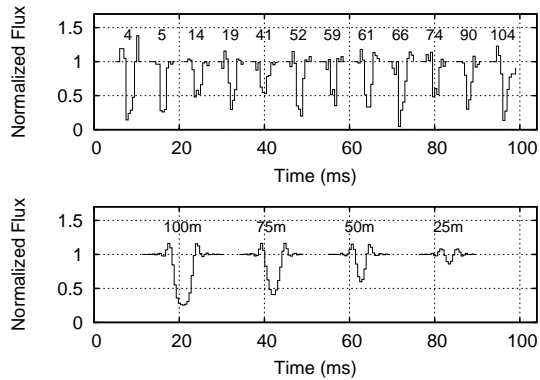}
  \caption[Reproductions of the 12 candidate occultation events (C07)
  compared to model lightcurves]
  {Reproductions of the 12 candidate occultation events (C07) compared
    to model lightcurves.  The events were manually reproduced from
    Figure 12 of C07, and event numbers are those assigned by C07.
    The model events simulate Sco X-1 occultations of 0.2-0.6nm x-rays
    by TNOs at 43 AU having diameters of: 25m, 50m, 75m, and 100m.}
  \label{fig:xraymodel}
\end{figure}

The x-ray occultation candidates reported by C07\footnote{Values for
  the flux levels were taken from \citet{chang07}, Figure 12} were
compared to model events (see Figure~\ref{fig:xraymodel}).  The Sco
X-1 x-ray source was modeled with $\lambda$=0.2-0.6nm light \citep[6-2
keV,][]{bradshaw03}, produced by a circular disk with a diameter of 3m
projected at 43 AU (C06).  Occultation shadows were made for 25m, 50m,
75m, and 100m KBOs at 43 AU.  The x-ray ($\lambda$=0.4nm) Fresnel
scale at 43 AU is 35m and these shadows are expected to be dominated
by diffraction effects.  The 25m and 50m models are of order
$\lesssim$1 Fsu and will produce circularly-symmetric diffraction
shadows, regardless of the projected shape of the occulter.  The 75m
and 100m models correspond to 2, and 3 Fsu.  Real (irregularly shaped)
objects this large would produce a more complex pattern of fringes,
but the overall width (duration) and depth of the lightcurves would be
comparable to those modeled here with a circular mask (the symmetry of
the shadows, as it relates to the size of the occulter, is further
discussed in Section~\ref{sec:discussion}). A relative velocity of
25$\kmps$ (consistent with a Keplerian orbit at 43 AU) was
used\footnote{Relative velocities were not provided by C07.}.

The 12 reported events range in duration from 1.5ms to 2.0ms (3 or 4
low flux measurements), and exhibit sharp transitions when dropping
from the baseline intensity and when returning to it.  Sections
through each of our modeled shadows were taken at an impact parameter
that maximized the depth-to-width ratio of the diffraction shadow
(b=40m). Non-zero impact parameters also avoided the flux increase
associated with the central flash\footnote{For occulters smaller than
  1 Fsu (or for a circular occulting mask of any size), the intensity
  at the center of the diffraction shadow is as though the occulter
  were not present.  The feature, called a `central flash', is due to
  constructive interference.} which was not observed in any of the C07
events.

Lightcurves for the 100m and 75m diameter TNOs are too broad (long in
duration) to be consistent with the observed events, and the 25m
object produces intensity changes which are too shallow.  The 50m
object's model lightcurve comes closest to reproducing the the
observed flux changes.  However, events with only 3 or 4 low flux
measurements are too narrow to be consistent with occultations by KBOs
(C07 event numbers 4, 5, 14, 19, 52, 59, 61, 66, 74, 90, and 104),
particularly in cases where 3 flux measurements are below $\sim$40\%
of the original intensity (C07 events 4, 5, 52, and 66).  The abrupt
decrease in flux observed in the remaining event (event 41) and many
others make them unconvincing when compared to the model lightcurves.

The absence of the central flash in {\itshape all} events is
statistically unusual.  For objects of the purported size, the center
of the shadow should display some rise in brightness, regardless of
any irregularities in the shape of the occulter.


As dozens of similar events were found to be instrument artifacts,
acceptance of these detections as TNO occultations appears premature.

\subsection{Three TNOs detected by \citet{roques06}}

\citet{roques06} (henceforth R06) reported 3 TNO occultation events
found in 604 minutes of 45 Hz imaging (simultaneous observations with
Sloan g$^\prime$ and i$^\prime$ filters).  The reported objects had
radii and distances of: 110m at 15AU, 300m at 210AU, and 320m at
140AU, and would be beyond the detectable limits of our observations.
There is no disagreement with our null result.

The sizes of the occulters present a serious inconsistency when
compared to the C06 result.  With $\sim$10 hours of data, sensitive to
objects with impact parameters we estimate would be comparable to our
own ($b\approx$1500m, due to diffraction), R06 found two $\sim$600m
objects.  The $\sim$89 hours of X-ray data from C07 would be sensitive
to 600m objects with impact parameters of $b\approx$300m (no x-ray
diffraction effects for an object this large), and would be expected
to yield $\sim(bvt)_{C06}/(bvt)_{R06}$ = 1.8$\times$ as many 600m
objects -- $\sim$4.  C07 reported no flux drop-outs consistent with
objects this large.  Though we found the x-ray flux drop-outs to be
unconvincing as occultations by 50m TNOs, events produced by 600m TNOs
(had they been present) are unlikely to have been overlooked by C06/C07.
Consider that at 140-210 AU, the x-ray ($\lambda$=0.4nm) Fresnel scale
is 65-79m - much smaller than the purported objects.  Occultations
would have produced near complete attenuation of the x-ray source (Sco
X-1).  At $\sim$25 km/s relative velocity, a 600m object would occult
the source for 24ms and would occupy 48 consecutive photometric
samples (C07 used 0.5ms sampling).  Even an off-center chord through
such a shadow would have produced an obvious occultation which would
easily have been identified by the authors.  The absence of any
corroborating 600m TNOs in the x-ray time series is a serious
inconsistency between R06 and C07.

R06 use a `variability index' (VI) for their detection algorithm.  The
standard deviation was computed in 0.08-0.40s intervals through the
time series (each denoted `sig(int)'), and normalized by the number of
standard deviations (stddev(sig)) from the overall mean standard
deviation (meansig); thus: \mbox{VI=(sig(int)-meansig)/stddev(sig)}.
Localized increases above VI=4.5 were interpreted as being due to the
brightness fluctuations associated with occultation candidates.  Three
events were identified with VI=5.3, 5.6, and 7.2.  These are the
moduli of the vector (VI$_{g^\prime}$, VI$_{i^\prime}$), but as the
VI$_{i^\prime}$ values are small\footnote{The events do not appear as
  strongly in the i$^\prime$ time series.  This is expected as a
  result of the wavelength dependence of diffraction.}, the modulus of
the vector is well approximated by the VI$_{g^\prime}$ value.

The VI values were interpreted as (VI)-sigma results, but the
underlying distribution of standard deviations (sig(int) values) used
to compute the VI values is non-Gaussian; it is positively skewed.

\begin{figure}[htbp]
  \plotone{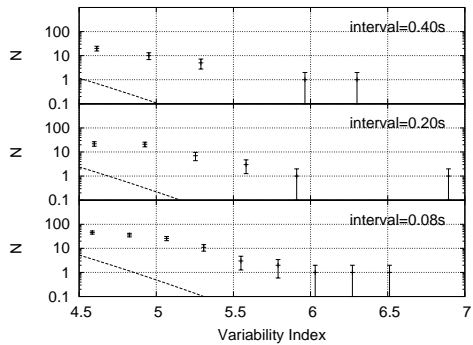}
  \caption[The distribution of VI values expected for 1/f noise]
  {The distribution of VI values expected for 1/f noise.  VI values
    were computed for intervals of 0.08s (bottom), 0.20s (middle), and
    0.40s (top); for a 10 hour (45 Hz) time series of 1.5\% 1/f noise.
    Dashed lines indicate the levels for Gaussian variates.  High VI
    values occur more frequently than expected for Gaussian variates.} 
  \label{fig:VIhist}
\end{figure}

To determine the VI values expected in the data, a 10 hour (45 Hz)
time series with 1.5\% 1/f noise was artificially generated
(guaranteed to contain no occultations).  The VI was computed with
intervals of 0.08s, 0.20s, and 0.40, consistent with those described
in R06.  For each interval, the distribution of values larger than
VI=4.5 (the R06 threshold) are shown in Figure~\ref{fig:VIhist}.  The
number of VI$>$4.5 values is higher than expected for Gaussian
variates.  Values as high as those observed by R06 are expected with
this method.  Values as high as VI=7.2 were not observed, but a
temporary increase in the noise level of a time series could easily
produce such a measurement.

In this evaluation, several tens of points were found with VI$>$4.5.
This behavior is not reported by R06.  However, the method produces a
positively skewed distribution of variates which can be seen R06
Figure 1.  The best that can be stated at this time is that the
statistical significance of the points is lower than would be the case
for Gaussian variates, but is otherwise unclear.

\section{The Surface Density  Limit}
\label{sec:upperlimit}

The slope of the size-distribution for the sub-km objects is estimated
to be q$\simeq$3.5.  A variety of independent methods have been used,
including: numerical simulations of the collisional evolution
\citep{kenyon04,pan05}, observations of cratering on Triton
\citep{stern00}, and inferences based on the absence of optical and
infrared light produced by such a population \citep{kenyon01}.

Although we are skeptical of the reported sub-km TNO occultations for
the reasons stated in Section~\ref{sec:critique},
neither C06, nor R06, nor this work have reported any occultation by a
TNO larger than 1 km.  At this limiting size, an upper limit
more stringent than our own can be placed on the current sky surface
density of TNOs.

Equation~\ref{eq:densitylimit} was recalculated with the $bvt$
products for each data set summed.  Velocity and impact parameter
values of ($v=23\kmps$,$b=1.5$km) and ($v=25\kmps$,$b=500$m) were used
for the R06 and C06 data set, respectively.  The velocities are the
average relative velocities for KBOs during the observation
periods.  The impact parameter for R06 (b=1.5km) was chosen to
accommodate diffraction effects.  Such effects would be negligible for
an x-ray occultation by a 1km object.

\begin{figure}[htbp]
  \plotone{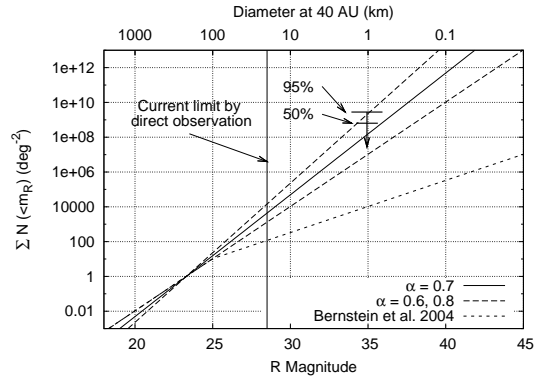}
  \caption[The CLF showing the overall upper limit on 1 km KBOs]
  {The CLF showing the overall upper limit on 1km KBOs.  Upper limits
    for 95\% and 50\% confidence are shown with horizontal lines above
    a downward pointing arrow.  Slopes of $\alpha$=0.6, 0.7, and 0.8
    are shown with the model CLF of \citet{bernstein04}.  A vertical
    line is used to indicate the current limit for direct
    observation \citep{bernstein04}.}
  \label{fig:clftotal}
\end{figure}

Thus, the sum of $bvt$ over each target is \mbox{$\sum_i b_i v_i t_i =
5.4\times 10^{-10}$ deg$^{2}$}, yielding an overall 95\% confidence limit
on the surface density of 1 km (diameter) KBOs at 40 AU of \mbox{$\sum
  N (m_R<35.0)$} = \mbox{$2.8\times 10^9$ deg$^{-2}$}.  This limit is shown in
Figure~\ref{fig:clftotal}. 

\section{Other Considerations}
\label{sec:discussion}

\begin{figure}[htbp]
  \plotone{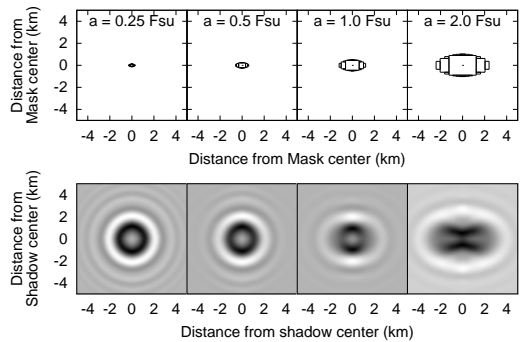}
  \caption[Occulting masks and diffraction shadows for non-spherical KBOs.]
  {Occulting masks (upper panels) and diffraction shadows (lower
    panels) for different-sized non-spherical KBOs (a/b=2.5 in each
    case, where a and b are the KBO's semi-major and semi-minor axes.)
    The diffraction shadows are circularly symmetric for a$\lesssim$0.5
    Fsu ($\sim$650m for visible light at 40 AU), despite the asymmetry
    of the occulting masks.  The grayscales were chosen individually
    to cover the range of each shadow's intensity.}
  \label{fig:symmetry}
\end{figure}

Non-spherical objects originating in the outer solar system (eg.
peanut-shaped comets) suggest that the spherical symmetry assumed in
our lightcurve models is not to be expected for TNOs.  However,
diffraction shadows cast by objects with R$_{KBO}\lesssim$0.5 Fsu are
circularly symmetric, regardless of the shape of the occulter.
Occultation shadows for different-sized elliptical TNOs ($a/b = 2.5$,
where $a/b$ is the ratio of the semi-major and semi-minor axes) were
generated to illustrate this (see Figure~\ref{fig:symmetry}).  The
circular symmetry of the shadows is preserved for $a_{KBO}\lesssim$0.5
Fsu ($\sim$650m for visible light at 40 AU), despite the asymmetry of
the occulting masks.  The shadow profiles for objects larger than this
do not require cross-correlation to detect; they would have been
identifiable by visual inspection of the time series, had any been
present.

\begin{figure}[htbp]
  \plotone{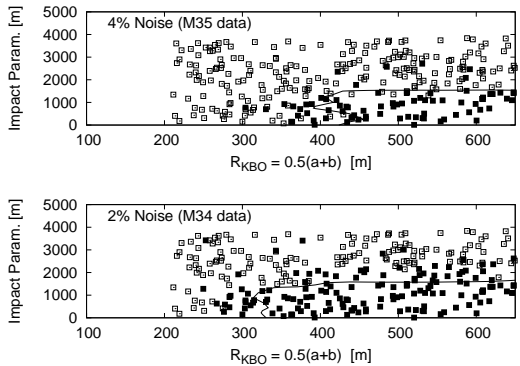}
  \caption[The recovered sizes and impact parameters of planted
  occultation events for elliptical objects] 
  {The recovered sizes and impact parameters of planted occultation
    events for elliptical objects.  Squares represent planted
    occultation events; filled if recovered, and open if not
    recovered.  The 80\% contour from the theoretical recovery tests
    for {\itshape circular} 40AU objects
    (Section~\ref{ssec:efficiency}) is shown with a solid line.  The
    planted events had parameters of: 200m$<a<$1000m,
    0m$<$impact-parameter$<$4000m, $1.0<(a/b)<3.0$, and
    $0<\theta<\pi$; where $a$, $b$, and $\theta$ are the semi-major and
    semi-minor axes of the object, and $\theta$ is the position angle
    of the rotated object. R$_{KBO}$ was taken to be $0.5(a+b)$.}
  \label{fig:ellip_recovery}
\end{figure}

Plant and recover tests were performed with profiles for 500
elliptical objects having uniformly distributed random parameters of:
200m$<a<$1000m, 0m$<$impact-parameter$<$4000m, 1.0$<(a/b)<$3.0, and
$0<\theta<\pi$; where $a$, $b$, and $\theta$ are the semi-major and
semi-minor axes of the object, and $\theta$ is the position angle of
the rotated object. R$_{KBO}$ was taken to be $0.5(a+b)$ in order to
compare the results to Figure~\ref{fig:recovered_kbos}. The observed
detection limits did not change and are shown in
Figure~\ref{fig:ellip_recovery}. Thus, cross-correlation detection
kernels based on circularly-symmetric occulting masks successfully
detect occultations by sub-Fresnel-scale irregularly-shaped KBOs.

Our method of generating artificial time series did not incorporate
the effects of variable noise levels in a time series.  Although many
of our recovery tests were performed with constant-noise artificial
time series, they were verified in our final detection run by
planting artificial occultation events in the real data.  The results
were found to be consistent with simulations we performed with the
artificial data, and we believe that the presented estimates for the
limiting detectable KBO size, $r_{min}$, and the maximum
impact parameter, $b_{max}$, are accurate.

The presence of a period of increased noise within a time series
alters the statistical properties of the variates.  The overall
{\itshape measured} standard deviation of the variates will lie
between those measured for the low-noise and high-noise segments,
which will lead to increases in the rates of false negatives and false
positives in the respective segments.  Our time series were broken
into segments having approximately constant noise levels to mitigate
this problem.

\section{Summary}
\label{sec:summary}

Analysis of our 5.0 star-hours of photometric time series on two B9V
stars in the ecliptic open cluster, M35, revealed no viable candidate
occultations.  For the first time, the recovery of artificial
occultation lightcurves was used to determine the limiting detectable
object size and impact parameter for a KBO occultation.  We place an
upper limit on the surface density of KBOs of \mbox{$\sum N (m_R<35.3)$} =
\mbox{$3.5\times 10^{10}$ deg$^{-2}$} (m$_R\simeq$35.3 corresponds to a
$\sim$860m (diameter) object with a geometric albedo of 0.04).

Having evaluated positive detections of TNO occultation events
\citep{chang06,chang07,roques06}, we believe that the published events
are likely to be spurious and that no serendipitous stellar
occultation by a TNO has yet been reported.  Comparison of the
expected lightcurve structure for the C07 events with the observed
structures show that the decreases in flux are too brief to be
consistent with occultations by small bodies at 40 AU; and evaluation
of the variability index, VI, used by R06 revealed that the {\itshape
  VI variates} are not normally distributed, but are drawn from a
positively skewed distribution, prone to yielding high values.  Also,
the two $\sim$600m (diameter) TNOs discovered in 10 star-hours of R06
data suggest $\sim$4 such events should be present in the 89
star-hour x-ray time series observed at a similar ecliptic latitude
($\beta_{R06}=0.4\arcdeg S,\ 0.4\arcdeg S,\ 7.1\arcdeg N$,
$\beta_{C06}=6\arcdeg N$).  Nothing this large was reported by C06/C07.

Combining all results from the different surveys we place an overall
95\%-confidence upper limit on the TNO surface density of: \mbox{$\sum
  N (m_R<35.0)$} = \mbox{$2.8\times 10^{9}$ deg$^{-2}$}.  A direct
extrapolation of the observed CLF slope ($\alpha\approx$0.7) would
yield a density of \mbox{$\sum N (m_R<35.0)$} = \mbox{$1.9\times
  10^{8}$ deg$^{-2}$}.

Our work is the first complete and careful examination of
observational data for KBO occultations and their proper analysis and
interpretation.


\acknowledgments The authors wish to thank the Herzberg Institute of
Astrophysics (HIA) for its support of the project through the Morton
Fellowship, and we wish to extent special thanks to Tim Hardy, Les
Saddlemyer, and Marc Baril at HIA for their work in the development of
the high-speed camera which was used to perform our observations.
This research used the facilities of the Canadian Astronomy Data
Centre (CADC) operated by the National Research Council (NRC) of
Canada with the support of the Canadian Space Agency (CSA), and was
supported by a Discovery Grants to DLW and JJK by the Natural Sciences
and Engineering Research Council of Canada (NSERC).  We are also
grateful to the Fund for Astrophysical Research (FAR) for their
financial support in the development of the high-speed camera
(\url{foundationcenter.org/grantmaker/fundastro/}).

{\it Facilities:} \facility{DAO}, \facility{HIA}, \facility{CADC}.


\end{document}